\documentclass[twocolumn,twocolappendix]{aastex631} 

\usepackage{amsmath,amssymb,amsfonts,bm}
\usepackage{epstopdf}
\usepackage{epsfig}
\usepackage{natbib,caption2}
\usepackage{graphicx}   
\usepackage{float}
\usepackage{longtable}
\usepackage{graphics}
\usepackage{hyperref}
\usepackage{color}
\usepackage{calc}
\usepackage{threeparttable}

\newcommand \beq{\begin{equation}}
\newcommand \eeq{\end{equation}}
\newcommand \bey{\begin{eqnarray}}
\newcommand \eey{\end{eqnarray}}

\newcommand \pc{\, {\rm pc} }

\newcommand \Msun{M_\odot} 
\newcommand \Msunyr {\rm M_\odot~yr{^{-1}}}
\newcommand \Msunpc {\rm M_\odot~pc{^{-2}}}

\newcommand \Ms{M_{\star}}

\newcommand{\Reff}{R_{\rm{e}}}
\newcommand{\Se}{\Sigma_{\rm{e}}}
\newcommand{\logSe}{\log\Sigma_{\rm{e}}}

\newcommand{\gsim}{\lower.5ex\hbox{$\; \buildrel > \over \sim \;$}}
\newcommand{\lsim}{\lower.5ex\hbox{$\; \buildrel < \over \sim \;$}}
\newcommand{\ha}{\hbox{H$\alpha$}}
\newcommand{\hb}{\hbox{H$\beta$}}

\accepted{July 22, 2022}
\acceptjournal{ApJL}

\shortauthors{Chen et al.}

\begin{document}

\title{Discovery of a Bimodal Environmental Distribution of Compact Ellipticals in the Local Universe}

\correspondingauthor{Hong-Xin Zhang, Xu Kong}
\email{guangwen@mail.ustc.edu.cn \\ hzhang18@ustc.edu.cn \\ xkong@ustc.edu.cn}

\author[0000-0002-4742-8800]{Guangwen Chen}
\affiliation{CAS Key Laboratory for Research in Galaxies and Cosmology, Department of Astronomy, University of Science and Technology of China, Hefei 230026, China}
\affiliation{School of Astronomy and Space Sciences, University of Science and Technology of China, Hefei, 230026, China}

\author[0000-0003-1632-2541]{Hong-Xin Zhang}
\affiliation{CAS Key Laboratory for Research in Galaxies and Cosmology, Department of Astronomy, University of Science and Technology of China, Hefei 230026, China}
\affiliation{School of Astronomy and Space Sciences, University of Science and Technology of China, Hefei, 230026, China}

\author[0000-0002-7660-2273]{Xu Kong}
\affiliation{CAS Key Laboratory for Research in Galaxies and Cosmology, Department of Astronomy, University of Science and Technology of China, Hefei 230026, China}
\affiliation{School of Astronomy and Space Sciences, University of Science and Technology of China, Hefei, 230026, China}
\affiliation{Frontiers Science Center for Planetary Exploration and Emerging Technologies, University of Science and Technology of China, Hefei, Anhui, 230026, China}

\author[0000-0001-8078-3428]{Zesen Lin}
\affiliation{CAS Key Laboratory for Research in Galaxies and Cosmology, Department of Astronomy, University of Science and Technology of China, Hefei 230026, China}
\affiliation{School of Astronomy and Space Sciences, University of Science and Technology of China, Hefei, 230026, China}

\author[0000-0002-2384-3436]{Zhixiong Liang}
\affiliation{CAS Key Laboratory for Research in Galaxies and Cosmology, Department of Astronomy, University of Science and Technology of China, Hefei 230026, China}
\affiliation{School of Astronomy and Space Sciences, University of Science and Technology of China, Hefei, 230026, China}

\author[0000-0002-2178-5471]{Zuyi Chen}
\affiliation{Steward Observatory, University of Arizona, 933 N Cherry Ave, Tucson, AZ 85721, USA}

\author[0000-0003-2876-577X]{Yimeng Tang}
\affiliation{CAS Key Laboratory for Research in Galaxies and Cosmology, Department of Astronomy, University of Science and Technology of China, Hefei 230026, China}
\affiliation{School of Astronomy and Space Sciences, University of Science and Technology of China, Hefei, 230026, China}
\affiliation{Department of Astronomy and Astrophysics, University of California, Santa Cruz, 1156 High Street, Santa Cruz, CA 95064, USA}

\author[0000-0002-5016-6901]{Xinkai Chen}
\affiliation{CAS Key Laboratory for Research in Galaxies and Cosmology, Department of Astronomy, University of Science and Technology of China, Hefei 230026, China}
\affiliation{School of Astronomy and Space Sciences, University of Science and Technology of China, Hefei, 230026, China}




\begin{abstract}

Low-mass compact stellar systems (CSSs; $M_{\star}$ $<$ 10$^{10}$ M$_{\odot}$) are thought to be a mixed bag of objects with various formation mechanisms. Previous surveys of CSSs were biased to relatively high-density environments and cannot provide a complete view of the environmental dependence of the formation of CSSs. We conduct the first-ever unbiased flux-limited census of nearby quiescent CSSs over a total sky area of $\sim$ 200 deg$^{2}$ observed by the GAMA spectroscopic survey. The complete sample includes 82 quiescent CSSs, of which 85\% fall within the stellar mass range of classical compact ellipticals (cEs).\ By quantifying the local environment with the normalized projected distance $D/R_{\rm vir}$ to the nearest luminous neighboring galaxy, we find that these CSSs have a bimodal $D/R_{\rm vir}$ distribution, with one group peaking near $\sim$ 0.1$\times$$R_{\rm vir}$ (satellite) and the other peaking near $\sim$ 10$\times$$R_{\rm vir}$ (field). In contrast to the CSSs, ordinary quiescent galaxies of similar masses have unimodal $D/R_{\rm vir}$ distribution.\ Satellite CSSs are older and more metal-rich than field CSSs on average. The bimodal $D/R_{\rm vir}$ distribution of quiescent CSSs reinforces the existence of two distinct formation channels (tidal stripping and born-to-be) for cEs and may be understood in two mutually inclusive perspectives, i.e., substantial tidal stripping happens only when satellite galaxies travel sufficiently close to their massive hosts, and there exists an excess of high-density cE-bearing subhalos close to massive halos.

\end{abstract}

\keywords{Compact dwarf galaxies (281); Compact galaxies (285); Elliptical galaxies (456); Galaxy formation (595); Stellar populations (1622); Tidal interaction (1699)}

\section{Introduction}

There exist a variety of compact stellar systems (CSSs) that lie in between classical globular clusters and ordinary galaxies in the size--luminosity plane ($10^6 \lesssim \Ms/\Msun \lesssim 10^{10}$; $10\lesssim\Reff \lesssim 600 \pc$).\ Well-known types of CSSs include super massive star clusters, ultracompact dwarfs (UCDs; \citealt{Hilker_1999, Drinkwater_2000, Brodie2011}), and compact ellipticals (cEs; \citealt{Chilingarian_2009Sci,Huxor_2013, Norris2014, Chilingarian_2015Sci, Kim_2020}).\

One prevailing scenario for the origin of these intermediate-mass CSSs is that they are the central remnants of tidally stripped nucleated galaxies. This scenario is motivated by the fact that many CSSs are spatially associated with more massive neighbor galaxies or lie in galaxy clusters/groups (e.g., \citealt{Chilingarian_2009Sci, Price2009}). Discoveries of tidal streams and over-massive central black holes in several CSSs support the tidal stripping scenario (e.g., \citealt{Seth_2014,Pechetti2022}).\ A series of circumstantial evidence for this scenario has been found for UCD system in the Virgo cluster \citep[][]{Zhang_2015, Liu_2015a}.\ UCDs follow mass-metallicity relations similar to nuclear star clusters, instead of ordinary globular clusters \citep{Zhang_2018}.\ Alternatively, CSSs, especially those at the low-mass end, may be simply super massive star clusters or their merger products \citep[e.g.,][]{Fellhauer_2002, Mieske2012}. Ordinary galaxies may evolve into compact galaxies through environmental regulation of star formation \citep{Du2019}. The discovery of apparently isolated cEs implies that they may be born to be compact galaxies (e.g., \citealt{Huxor_2013, Paudel_2014, Rey_2021}), and some of them may be merger remnants of smaller galaxies \citep[e.g.,][]{Paudel_2014}.\ Some group/cluster cEs may be ejected via a three-body interaction and become isolated \citep{Chilingarian_2015Sci}.

Given their intermediate nature, it is conceivable that CSSs constitute a mixed bag of various formation mechanisms that depend on properties such as mass and environment. A vast majority of previous studies focused on CSSs located in relatively high-density environments (e.g., galaxy groups, clusters). \citet{Chilingarian_2015Sci} and \citet{Kim_2020} are among the few studies that selected large samples of cEs (195 and 138, respectively) over a large sky area that covers different environments, thanks to the Sloan Digital Sky Survey (SDSS; \citealt{York_2000}).\ 



Previous studies lack a complete (either in a volume-limited or flux-limited sense) and large-sky-area (covering a variety of environments) census of CSSs that allows for an unbiased investigation of the formation of CSSs.\ This is because most large-scale spectroscopic surveys such as SDSS suffer from the fiber-collision problem, which severely affects the census of CSSs on small scales. The Galaxy and Mass Assembly spectroscopic survey (GAMA; \citealt{Driver_2011_GAMA}) largely overcomes the fiber-collision problem by adopting a multipass observing strategy. Moreover, GAMA significantly improves the completeness level of compact galaxies \citep{Baldry_2010}. In this work, we perform the first flux-limited census of nearby CSSs based on the GAMA survey and study the structural properties and stellar populations of CSSs over different local environments.\ Throughout this work, we assume a flat $\Lambda$CDM cosmology with $\Omega_m = 0.3$, $\Omega_\Lambda = 0.7$, and $h_0 = 0.7$.\ 

\section{Sample selection} \label{sec:data}

\subsection{Compact Stellar Systems}  \label{sec:sample}

The latest data release of GAMA (DR4; \citealt{GAMA_DR4}) covers three equatorial regions (field IDs: G09, G12, G15) and two southern regions (field IDs: G02, G23). The equatorial regions are 95\% complete down to $r \simeq 19.7$ mag. The G02 and G23 fields are complete down to $r \simeq 19.8$ and 19.4 mag respectively.\ The GAMA collaboration did not observe galaxy candidates that meet their selection criteria but had already been observed by pre-existing spectroscopic surveys (e.g., SDSS). So a complete spectroscopic sample (as used in this work) in the GAMA sky area includes those observed by both the GAMA team and other pre-existing surveys (\citealt{Driver_2011_GAMA}). 

To measure the size of CSSs that are usually marginally-resolved in ground-based images, we turn to the Hyper Suprime-Cam Subaru Strategic Program (HSC-SSP; \citealt{HSC_Overview}) wide layer imaging survey which surpasses previous surveys in its combination of sky coverage and imaging quality (e.g., $i$-band seeing $\sim$ 0.6$''$; $r_{\rm limit}$ $\simeq$ 26 mag). The GAMA fields G09, G12, G15, and part of G02 fall within the sky coverage of the latest data release of HSC-SSP (DR3; \citealt{HSC_DR3}), with an overlapping area of $\sim$ 200 ${\rm deg^2}$ in total. Our CSS sample is selected from this overlapping sky area of GAMA and HSC-SSP (hereafter GAMA-HSC). 

\begin{figure}[!ht]
  \centering
  \includegraphics[width=0.47\textwidth]{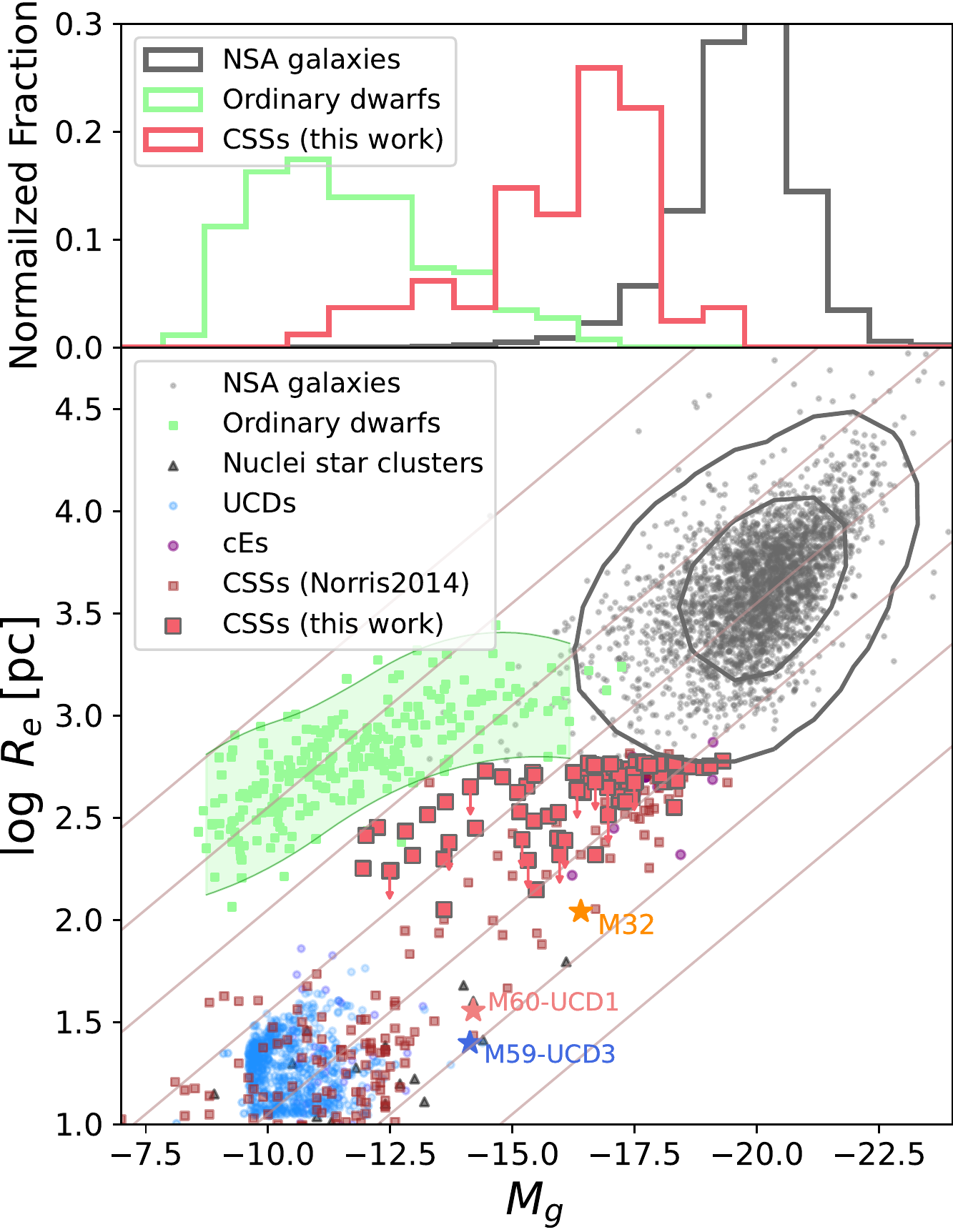} 
  \caption{
  Upper panel: $g$-band absolute magnitude distributions for galaxies from the NASA Sloan Atlas catalog (NSA; \citealt{Blanton2011}) within the same redshift range as our sample, for ordinary dwarf galaxies from \cite{Eigenthaler_2018}, and for CSSs in this work, are plotted as gray, green and red histograms, respectively.\
  Lower panel: $\Reff$--$M_g$ distribution. Red squares represent our CSSs and symbols with arrows attached indicate unresolved CSSs. The black contours enclose the central $68\%$ and $95\%$ of NSA galaxies. Only 3000 randomly chosen galaxies in the NSA catalog are plotted (gray dots). The green shaded region encloses $95\%$ of ordinary dwarf galaxies.
  Nuclear star clusters \citep{Misgeld2011,Brodie2011} are plotted as filled triangles.
  UCDs \citep{Zhang_2015,Liu_2020,Saifollahi2021} and cEs \citep{Huxor_2013,Paudel_2014,Guerou_2015} are plotted as blue and purple dots, respectively.
  CSSs from \cite{Norris2014} are plotted as small brown squares. M32, M60-UCD1 (\citealt{Strader_2013}) and M59-UCD3 (\citealt{Liu_2015b}) are plotted as filled stars. The diagonal lines represent constant $g$-band surface brightness.
  }
  \label{Fig:Size-M}
\end{figure}

The selection procedure starts from all GAMA-HSC objects with heliocentric velocities $>$ 500 km/s and redshift $< 0.1$\footnote{Only objects with redshift quality flag $nQ\ge 3$ are considered for our analysis, following \cite{Liske_2015}.}.\ Then, we measure the effective radii $\Reff$ of these objects with GALFIT (\citealt{Peng_2010}) by adopting single S\'{e}rsic models. The median of the $\Reff$ measured in all available bands is used in our analysis.\ Our experiments suggest that reliable $\Reff$ can be measured with GALFIT down to 1/3 $\times$ the Full Width at Half Maximum (FWHM) of Point Spread Function (PSF) within 20\% uncertainties (See Appendix \ref{Sec:Appendix} for more details).\ For the unresolved objects (i.e., $\Reff < $ 1/3 $\times$ the FWHM of PSF), we conservatively adopt half of the FWHM of PSF as the upper limit of $\Reff$. We select objects with $\Reff < 600 \pc$, following previous searches of CSSs (e.g., \citealt{Norris2014, Chilingarian_2015Sci, Kim_2020, Rey_2021}). Next, we exclude objects that fall above the lower boundary of the 95\% range of the size-magnitude distribution of ordinary dwarf galaxies, as illustrated by the green shaded region in Figure \ref{Fig:Size-M}. Lastly, we carry out a visual inspection of images of the above-selected CSS candidates to exclude H {\sc ii} regions that are part of other galaxies. 

The above selection procedure results in a sample of 395 CSSs candidates. By using a demarcation at $\log {\rm sSFR} = -11$, where ${\rm sSFR} \equiv{\rm SFR/\Ms}$ (see Section \ref{sec:spm}), we divide the sample into 313 star-forming and 82 quiescent CSSs, 12 of which are unresolved. None of our CSSs exhibits broad emission lines that would be a signature of Type 1 active galactic nuclei (AGN). The smallest heliocentric velocity of this sample turns out to be 887 km/s.\ We will focus on the 82 quiescent CSSs (hereafter CSSs for brevity) in the rest of the paper.\ Figure \ref{Fig:Size-M} shows the size--magnitude distribution of our CSSs (red data points).\ 

It is worth noting that any large-scale spectroscopic galaxy surveys inevitably miss some of the most compact galaxies (with respect to the spatial resolution) in order to maintain a reasonable survey efficiency.\ Nevertheless, GAMA makes an effort to significantly alleviate this potential issue by relaxing the star-galaxy separation criteria for the input catalog of photometric candidates. According to \cite{Baldry_2010}, GAMA is nearly complete at $\Delta_{sg} > 0.05$ mag, where $\Delta_{sg}$ is the PSF minus model magnitude difference.\ For comparison, the completeness limit of the SDSS main survey is $\Delta_{sg} > 0.24$ mag.\ It is not straightforward to estimate the fraction of potentially missed CSSs due to the $\Delta_{sg}$ limit imposed by GAMA. However, we point out that our conclusion in this paper would not change if raising the $\Delta_{sg}$ limit, for instance, to $\Delta_{sg} > 0.24$ (including 63, or 77\%, of our CSSs).

\subsection{Control sample of ordinary galaxies}

We also select a control sample of ordinary quiescent galaxies from the same sky fields of GAMA.\ For each CSS in our sample, we randomly select an ordinary galaxy that has $\log  {\rm SFR}$ and $\log\Ms$ that are each within $\pm$0.3 dex of the CSS. Here the ``ordinary'' galaxies are defined to be above the upper boundary of our CSS sample on the $\Reff$-$M_{g}$ plane.\ We have repeated the random sampling of ordinary galaxies for several tens of times, and found that the relevant conclusion for the control sample in this work does not vary. So the following analysis will be based on one of these randomly selected control sample.

\section{Analysis}

\subsection{Quantifying the Local Environment} \label{sec:envpar}

To quantify the local environment, we derive $\log D/R_{\rm vir}$ for both the CSS and control samples, where $D$ is the projected distance to the nearest luminous neighboring galaxy that is at least 2.0 mag more luminous and has a radial velocity within $\pm 500$ ${\rm km}$ ${\rm s^{-1}}$ of the galaxy in question, and $R_{\rm vir}$ is the virial radius of the selected luminous neighbor, approximated to be 67 times its effective radius (\citealt{Kravtsov_2013}).\ We note that $\log D/R_{\rm vir}$ was derived by requiring the nearest luminous neighbor to have $M_r < -21$ mag in \cite{Kim_2020}. We also calculate $\log D/R_{\rm vir}$ for our samples in the same way as \cite{Kim_2020} in order to check the influence of different definitions on our results. As will become clear in Section \ref{sec:environment}, the sample can be divided into field and satellite subsamples, with a demarcation at $D/R_{\rm vir}$ $\simeq$ 0.4.

We also derive the projected galaxy number density parameter: $\eta_k = \log(\frac{3(k-1)}{4\pi d^3_k})$ \citep{Argudo_2015}, where $d_k$ is the projected distance to the 5th (i.e., $k$ = 5) nearest galaxy in Mpc.\ Only neighbors with $M_r < -18.6$ are considered in $\eta_{\rm k}$ calculation, corresponding to the GAMA completeness limit (i.e., $r$ = 19.72 mag) at the distance of the most distant CSS in our sample.

Lastly, we mention that 22 of the 82 CSSs belong to groups with at least 10 member galaxies according to the GAMA galaxy group catalogue (G3C; \citealt{GAMA_G3C_2011}).

\subsection{Stellar population modeling} \label{sec:spm}

Stellar masses of our CSSs are estimated by fitting broadband spectral energy distribution with the Code Investigating GALaxies Emission (CIGALE; \citealt{CIGALE_2019}), using the optical photometry from SDSS and HSC-SSP, and near-infrared photometry from the Wide-field Infrared Survey Explorer (WISE; \citealt{Wright_2010}).\ Star formation rates (SFR) are estimated from $\ha$ luminosities offered by GAMA DR4, by adopting the formula, ${\rm \log SFR [\Msunyr]} = \log L(\ha) {\rm [ergs}$ ${\rm s^{-1}]} -41.27$ \citep{Kennicutt_2012}, where $\ha$ is corrected for internal dust extinction from Balmer decrement by assuming an intrinsic flux ratio of $(\ha/\hb)=2.86$.\

Luminosity-weighted stellar ages and metallicities [$Z/$H] are estimated by performing a joint fitting of the (Gaussian) emission lines (if any) and spectral continuum (3540\AA~-- 5700\AA) of the spectra with the Penalized PiXel-Fitting (pPXF; \citealt{Cappellari_ppxf2017}) package, by using the MILES stellar library and non-parametric star formation histories. The individual spectrum was corrected for the Galactic extinction \citep{Schlegel1998} and shifted to the rest frame before the fitting.

\section{Results} \label{sec:results}



\subsection{Environmental Distribution} \label{sec:environment}

\begin{figure*}[!ht]
  \begin{center}
    \includegraphics[width=1.0\textwidth]{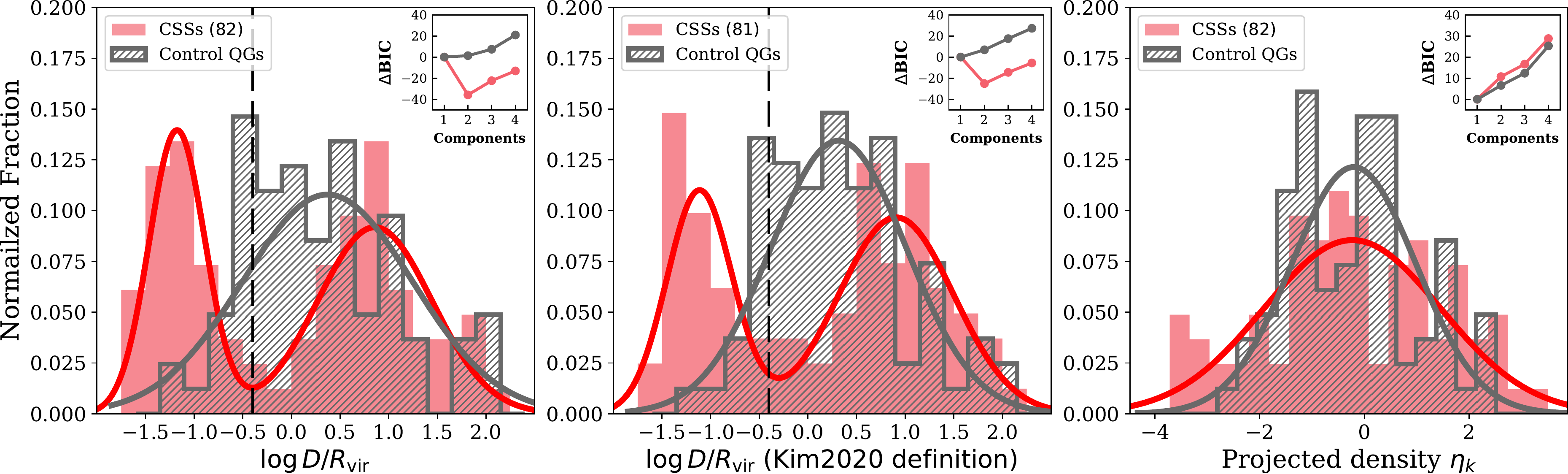}
    \caption{Normalized distribution of the environmental parameters of CSSs (red filled histogram) and control sample (grey hatched histogram). The left panel is for the projected distances to the nearest luminous neighboring galaxy selected with the default method described in the text and the middle panel is for the projected distances to the nearest luminous neighboring galaxy selected with the method in \cite{Kim_2020}. The projected distances have been normalized by the virial radii of the selected nearest neighbors. The right panel shows the distributions of the projected number density of nearby galaxies. The best-fit Gaussian curves from GMM fit to each sample are overplotted on the histograms. The inset figure in each panel shows the $\Delta \rm{BIC} \equiv \rm{BIC} \rm{(Component=n)} - \rm{BIC} \rm{(Component=1)}$ for GMM fit with different number n of components.
    See Section \ref{sec:envpar} for details of the environmental parameters plotted here.
    }
    \label{Fig:Dis_DtoR}
  \end{center}
\end{figure*}

Distribution of the environmental parameters of our CSSs and the control sample (Section \ref{sec:envpar}) are shown in Figures \ref{Fig:Dis_DtoR}.\ The CSSs appear to follow a bimodal $\log D/R_{\rm vir}$ distribution, with one group of CSSs peaking around $\log D/R_{\rm vir}$ $\sim$ $-1$ and the other group peaking around $\log D/R_{\rm vir}$ $\sim$ $1$. In contrast to the CSSs, the control sample appear to follow a unimodal distribution, which peaks at $\log D/R_{\rm vir}$ in between the two groups of CSSs.\ The projected number density $\eta_{\rm k}$ distribution appears to be unimodal for both samples. To verify this visual impression, we turn to the Python implementation of Gaussian Mixture Models (GMM) and use the Bayesian information criterion (BIC) to select the minimum number $N_{\rm comp}$ of Gaussian components that can adequately describe the observed distributions. 

The GMM fitting results are illustrated in Figure \ref{Fig:Dis_DtoR}.\ For CSSs, the BIC reaches minimum at $N_{\rm comp}$ = 2.\ With $\Delta$(BIC) $\simeq$ $-40$ (for our default luminous neighbor selection method; left panel) or $-30$ (for the \cite{Kim_2020} method; middle panel), the two-component models are strongly favored against one-component models.\ The two best-fit Gaussian subcomponents of the CSSs intersect at $\log D/R_{\rm vir}$ $\sim$ $- 0.4$. For the control sample, the BIC reaches minimum at $N_{\rm comp}$ = 1, verifying their unimodal distribution.\ The GMM fitting also verifies the visual impression of unimodal $\eta_{\rm k}$ distributions for both samples.

\subsection{Structural properties of CSSs and their hosts}\label{subsec:struct}

Given the above results, we split our CSSs into field and satellite subsamples by using $\log D/R_{\rm vir}$ = $-0.4$.

\begin{figure}[!ht]
    \centering
    \includegraphics[width=0.455\textwidth]{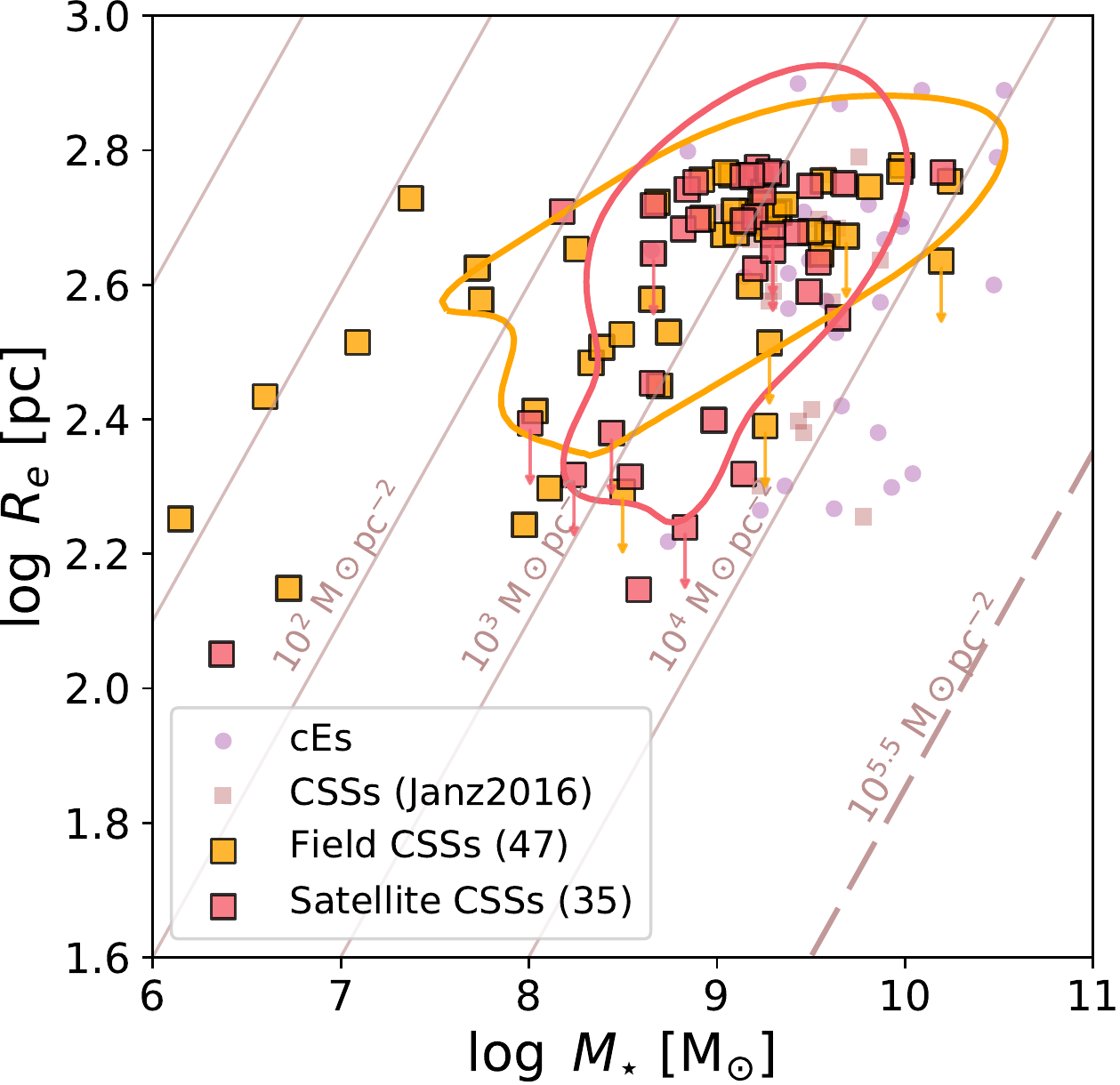}
    \caption{Size--mass distribution of our CSSs.\ Iso-number density contours that enclose $68\%$ of the field and satellite CSSs are overplotted with the same color scheme as the data points. The arrows indicate unresolved CSSs. Literature cEs \citep{Guerou_2015,Ferre_2018} and CSSs \citep{Janz_2016} are plotted as purple dots and brown squares, respectively.\ Lines of iso-$\Se \equiv \Ms/(2\pi\Reff^2)$ are indicated by brown slanted lines. The brown dashed line represents the maximum effective stellar mass density ($\Sigma_{\rm max} \simeq 10^{5.5} \Msunpc$) expected for CSSs in the local universe (\citealt{Hopkins2010}). 
    }
    \label{Fig:Dis_structure}
\end{figure}

The stellar size--mass distributions of the subsamples are exhibited in Figure \ref{Fig:Dis_structure}. Number density contours that enclose 68\% of the CSSs in each subsample are overplotted to guide the comparison between the subsamples. As indicated by the parallel lines of equal surface mass densities, the satellite CSSs cover a surface mass density range of $2 \lesssim \logSe [\Msun/\pc^{2}] \lesssim 4$ on the size-mass plane.\ Distribution of the field subsample is clustered at a similar surface mass density range, albeit with a tail extending to lower mass densities.\ 

To probe the morphologies of the nearest luminous neighboring galaxies of the satellite CSSs, we use the S\'{e}rsic indices ($n$) provided by the NSA catalog to classify the neighbors into early-type galaxies (ETGs: $n > 2.5$) and late-type galaxies (LTGs: $n < 2.5$). Not all of the neighboring galaxies have $n$ measurements. For those with $n$ measurements, 83\% (20) of the satellite CSSs have ETG neighbors.\ This finding is in line with the well-known ``galactic conformity'' phenomenon \citep{Weinmann2006}.

\begin{figure}[htp]
    \centering
    \includegraphics[width=0.47\textwidth]{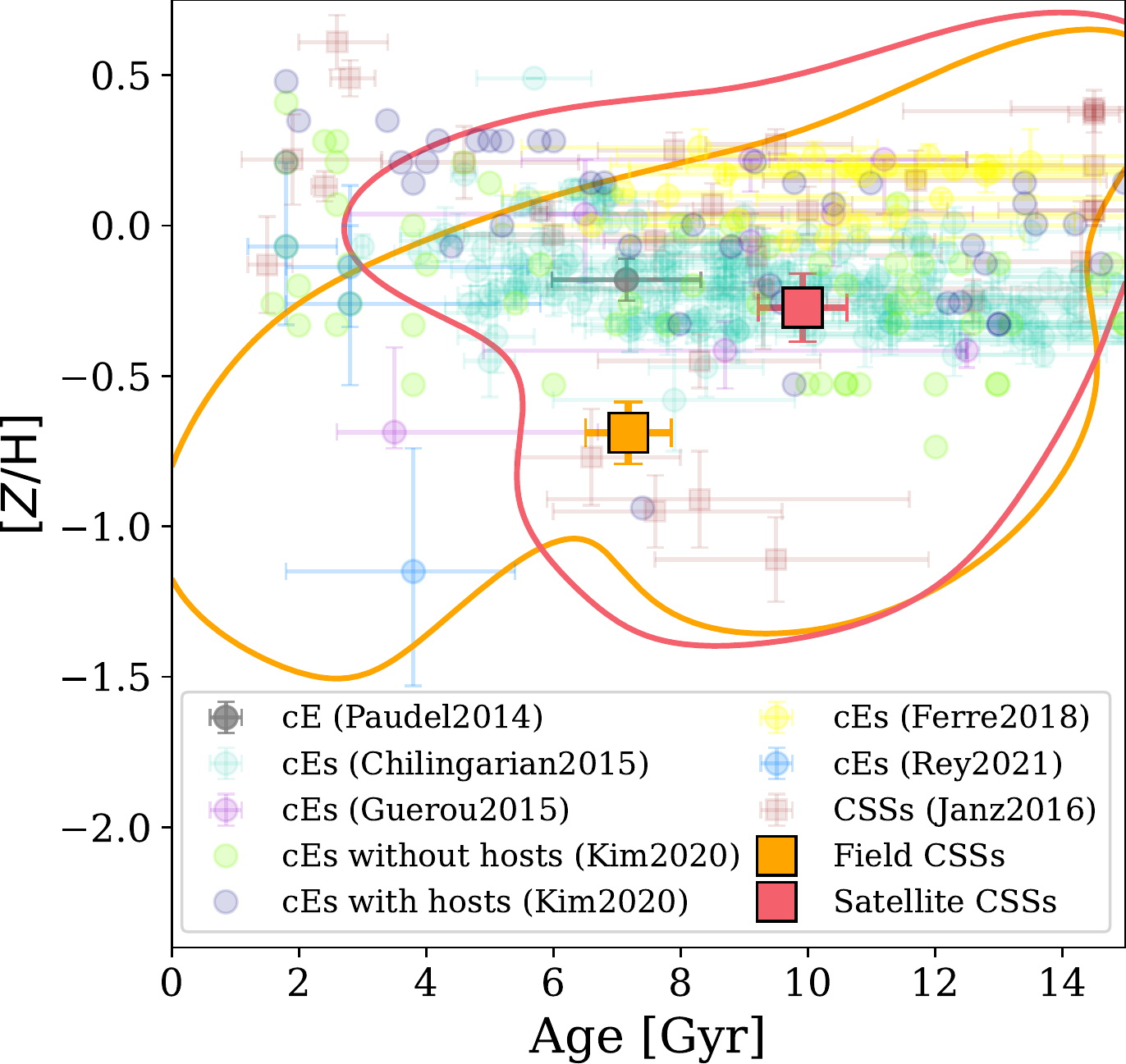}
    \caption{
    Age$-$[$Z/$H] diagram.\ The median values of the field and satellite subsamples of our CSSs are shown as big filled squares.\ The contour curves enclose 68\% of the CSSs in each of our subsample.\ Literature CSSs \citep{Janz_2016} or cEs \citep{Chilingarian_2015Sci,Guerou_2015,Ferre_2018,Kim_2020} are plotted with symbols indicated in the legend. The error bars of the median values of our CSSs are determined by random resampling of the original subsamples with replacement for 1,000 times.
    }
    \label{Fig:properties}
\end{figure}

\subsection{Stellar populations of CSSs}\label{subsec:stelpop}


Luminosity-weighted stellar ages and metallicities of the field and satellite subsamples of CSSs are shown separately in Figure \ref{Fig:properties}.\ Contours that enclose 68\% of the CSSs in each subsample and the median values are indicated in Figure \ref{Fig:properties}.\ Literature samples of cEs are also plotted for comparison purpose.\ Our CSSs cover nearly the same ranges of age and metallicities with the literature cEs.\ The satellite CSSs have significantly older median ages and higher median metallicities than the field CSSs. Systematically higher metallicities for satellite cEs were also reported by \cite{Kim_2020}.\

\section{Summary and Discussion} \label{sec:summary}

We have taken a complete (in a flux-limited sense) census of quiescent compact stellar systems (CSSs) over a total sky area of $\sim$ 200 deg$^{2}$ at $z<0.1$. The sky area was observed by the spectroscopic survey of GAMA and the wide-field imaging survey of HSC-SSP. GAMA largely overcomes the fiber-collision problem that has plagued the census of galaxies at small scales for most existing large-area spectroscopic surveys, and thus provides a unique opportunity to explore the environmental dependence of the formation of CSSs in an unbiased way. We selected 82 quiescent CSSs, and studied their small-scale environmental distribution, structural properties, and stellar populations. 

We quantify the local environment by using the normalized projected distance ($D/R_{\rm vir}$) to the nearest luminous neighboring galaxy, and find that these CSSs have a bimodal distribution, with one peak near $\sim$ 0.1$\times$$R_{\rm vir}$ and the other peak near $\sim$ 10$\times$$R_{\rm vir}$. Such bimodal distribution is not seen in ordinary quiescent galaxies.

We divide the CSSs into field and satellite subsamples based on their bimodal $D/R_{\rm vir}$ distribution, and find that the field and satellite CSSs follow similar mass-size distributions. Based on full-spectrum stellar population fitting, we find that the satellite CSSs are older and more metal-rich than field CSSs on average. A majority (83\%) of satellite CSSs have early-type hosts.\

The majority (85\%; Figure \ref{Fig:Dis_structure}) of our CSSs fall within the stellar mass range of cEs (10$^{8}-10^{10}$ M$_{\odot}$). A closely relevant study by \cite{Kim_2020} found an environmental dependence of the metallicity distribution of cEs, i.e., satellite cEs (especially those inhabiting rich groups) tend to have higher metallicities than field cEs of comparable masses, while field cEs largely follow the mass-metallicity relation of massive ordinary ETGs. A natural explanation for the different metallicity distributions is that, firstly, satellite cEs have experienced tidal stripping that significantly reduced the mass but barely affected the metallicities, and secondly, field cEs may be born to be the low mass extension of more massive ellipticals. Our finding of the bimodal $D/R_{\rm vir}$ distribution of quiescent CSSs reinforces the existence of two distinct formation channels of cEs \citep[e.g.,][]{Huxor_2011, Janz_2016, Ferre_2018}. 

The fact that a bimodal distribution instead of a skewed unimodal distribution is observed for the CSSs may be understood in two plausible perspectives. Firstly, in the tidal stripping framework, substantial tidal stripping happens only when satellites travel sufficiently close to the massive host \citep[e.g.,][]{Pfeffer2013, Mayes2021}; Secondly, in the hierarchical structure formation framework, CSSs inhabit higher-density subhalos than do the ordinary satellites and thus have a higher degree of central concentration toward massive halos \citep[e.g.,][]{Diemand2005}.

\section*{Acknowledgements}


This work is supported by the Strategic Priority Research Program of Chinese Academy of Sciences (No. XDB 41000000), the National Key R\&D Program of China (2017YFA0402600, 2017YFA0402702), the NSFC grant (Nos. 12122303, 11973038 and 11973039), and the science research grants from the China Manned Space Project (Nos. CMS-CSST-2021-A07, CMS-CSST-2021-B02).\ We also thank a support from the CAS Pioneer Hundred Talents Program.\ Z.S.L acknowledges the support from China Postdoctoral Science Foundation (2021M700137).

GAMA is a joint European-Australasian project based around a spectroscopic campaign using the Anglo-Australian Telescope. The GAMA input catalogue is based on data taken from the SDSS and the UKIRT Infrared Deep Sky Survey. Complementary imaging of the GAMA regions is being obtained by a number of independent survey programmes including GALEX MIS, VST KiDS, VISTA VIKING, WISE, Herschel-ATLAS, GMRT and ASKAP providing UV to radio coverage. GAMA is funded by the STFC (UK), the ARC (Australia), the AAO, and the participating institutions. The GAMA website is http://www.gama-survey.org/ .

This paper is based on data collected at the Subaru Telescope and retrieved from the HSC data archive system, which is operated by the Subaru Telescope and Astronomy Data Center (ADC) at NAOJ. Data analysis was in part carried out with the cooperation of Center for Computational Astrophysics (CfCA), NAOJ. 





\appendix

\section{Test of the Effective Radius Measurements}\label{Sec:Appendix}
  
\begin{figure}[!ht]
  \begin{center}
    \includegraphics[height=0.45\textheight]{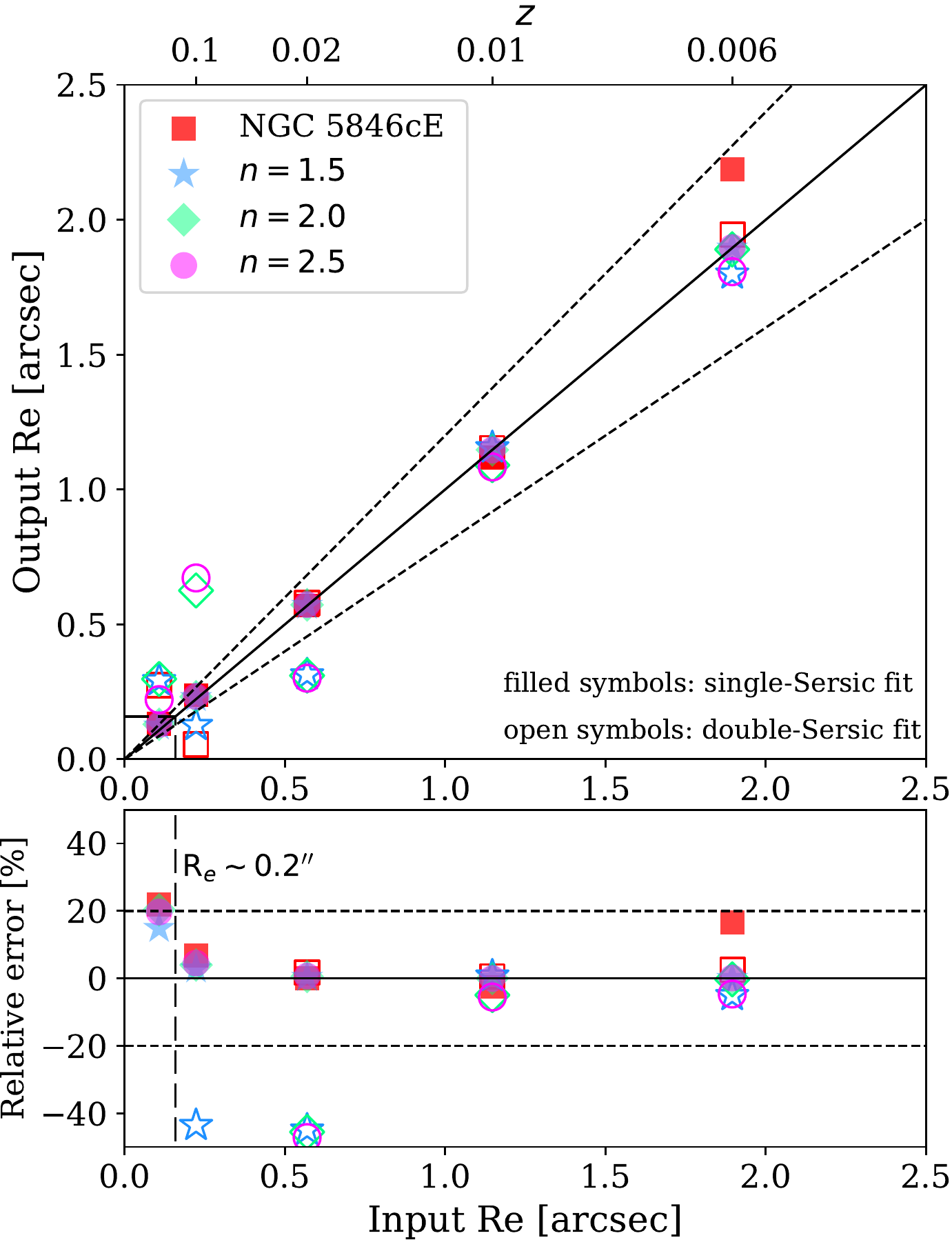}
    \caption{Comparison of the input and best-fit $R_e$ (by GALFIT) of mock galaxies.\ The squares correspond to mock galaxies with the same structural parameters with cE NGC5846cE, and the other symbols correspond to single-S\'{e}rsic mock galaxies with different S\'{e}rsic indices $n$.\ The same mock galaxies are put at a series of redshift, i.e., 0.1, 0.05, 0.02, 0.01, and 0.006.\ The filled (open) symbols correspond to GALFIT fitting with single (double) S\'{e}rsic functions.\ The vertical dashed lines indicate the $R_e$ measurement limit adopted in our work, which is about 1/3$\times$PSF FWHM.\ Note that a Gaussian distribution has $R_e$ $\simeq$ 1/2$\times$FWHM.}
    \label{Fig:Appendix}
  \end{center}
\end{figure}


We use mock galaxies to evaluate the limitation of HSC-SSP images for measuring the $\Reff$ of nearby CSSs.\ The test results are shown in Figure \ref{Fig:Appendix}.\ Specifically, we use the best-fit double-S\'{e}rsic parameters of the compact elliptical NGC5846cE (\citealt{Chilingarian_2010}) and a series of single-S\'{e}rsic profiles with different S\'{e}rsic indices to create PSF-convolved mock galaxy images for the test.\ The mock galaxies have intrinsic $\Reff$ = 240 pc and $V$-band absolute magnitude of $-17.07$ mag.\ Each of the mock galaxies is put at a range of redshifts and randomly placed at blank sky regions of the HSC-SSP images. We use GALFIT to fit these mock galaxies with either single (filled symbols in Figure \ref{Fig:Appendix}) or double (open symbols) S\'{e}rsic components, and compare the output $\Reff$ to the input in Figure \ref{Fig:Appendix}.\ Given this test, a single-S\'{e}rsic GALFIT fitting can recover $R_e$ to within 20\% at a limit of $R_e$ $>$ 1/3 PSF FWHM, below which the galaxies are regarded unresolved. 

\bibliography{cG}{}

\begin{thebibliography}{}
\expandafter\ifx\csname natexlab\endcsname\relax\def\natexlab#1{#1}\fi
\providecommand{\url}[1]{\href{#1}{#1}}
\providecommand{\dodoi}[1]{doi:~\href{http://doi.org/#1}{\nolinkurl{#1}}}
\providecommand{\doeprint}[1]{\href{http://ascl.net/#1}{\nolinkurl{http://ascl.net/#1}}}
\providecommand{\doarXiv}[1]{\href{https://arxiv.org/abs/#1}{\nolinkurl{https://arxiv.org/abs/#1}}}

\bibitem[{{Aihara} {et~al.}(2018){Aihara}, {Arimoto}, {Armstrong}, {Arnouts},
  {Bahcall}, {Bickerton}, {Bosch}, {Bundy}, {Capak}, {Chan}, {Chiba}, {Coupon},
  {Egami}, {Enoki}, {Finet}, {Fujimori}, {Fujimoto}, {Furusawa}, {Furusawa},
  {Goto}, {Goulding}, {Greco}, {Greene}, {Gunn}, {Hamana}, {Harikane},
  {Hashimoto}, {Hattori}, {Hayashi}, {Hayashi}, {He{\l}miniak}, {Higuchi},
  {Hikage}, {Ho}, {Hsieh}, {Huang}, {Huang}, {Ikeda}, {Imanishi}, {Inoue},
  {Iwasawa}, {Iwata}, {Jaelani}, {Jian}, {Kamata}, {Karoji}, {Kashikawa},
  {Katayama}, {Kawanomoto}, {Kayo}, {Koda}, {Koike}, {Kojima}, {Komiyama},
  {Konno}, {Koshida}, {Koyama}, {Kusakabe}, {Leauthaud}, {Lee}, {Lin}, {Lin},
  {Lupton}, {Mandelbaum}, {Matsuoka}, {Medezinski}, {Mineo}, {Miyama},
  {Miyatake}, {Miyazaki}, {Momose}, {More}, {More}, {Moritani}, {Moriya},
  {Morokuma}, {Mukae}, {Murata}, {Murayama}, {Nagao}, {Nakata}, {Niida},
  {Niikura}, {Nishizawa}, {Obuchi}, {Oguri}, {Oishi}, {Okabe}, {Okamoto},
  {Okura}, {Ono}, {Onodera}, {Onoue}, {Osato}, {Ouchi}, {Price}, {Pyo}, {Sako},
  {Sawicki}, {Shibuya}, {Shimasaku}, {Shimono}, {Shirasaki}, {Silverman},
  {Simet}, {Speagle}, {Spergel}, {Strauss}, {Sugahara}, {Sugiyama}, {Suto},
  {Suyu}, {Suzuki}, {Tait}, {Takada}, {Takata}, {Tamura}, {Tanaka}, {Tanaka},
  {Tanaka}, {Tanaka}, {Terai}, {Terashima}, {Toba}, {Tominaga}, {Toshikawa},
  {Turner}, {Uchida}, {Uchiyama}, {Umetsu}, {Uraguchi}, {Urata}, {Usuda},
  {Utsumi}, {Wang}, {Wang}, {Wong}, {Yabe}, {Yamada}, {Yamanoi}, {Yasuda},
  {Yeh}, {Yonehara}, \& {Yuma}}]{HSC_Overview}
{Aihara}, H., {Arimoto}, N., {Armstrong}, R., {et~al.} 2018, \pasj, 70, S4,
  \dodoi{10.1093/pasj/psx066}

\bibitem[{{Aihara} {et~al.}(2022){Aihara}, {AlSayyad}, {Ando}, {Armstrong},
  {Bosch}, {Egami}, {Furusawa}, {Furusawa}, {Harasawa}, {Harikane}, {Hsieh},
  {Ikeda}, {Ito}, {Iwata}, {Kodama}, {Koike}, {Kokubo}, {Komiyama}, {Li},
  {Liang}, {Lin}, {Lupton}, {Lust}, {MacArthur}, {Mawatari}, {Mineo},
  {Miyatake}, {Miyazaki}, {More}, {Morishima}, {Murayama}, {Nakajima},
  {Nakata}, {Nishizawa}, {Oguri}, {Okabe}, {Okura}, {Ono}, {Osato}, {Ouchi},
  {Pan}, {Plazas Malag{\'o}n}, {Price}, {Reed}, {Rykoff}, {Shibuya},
  {Simunovic}, {Strauss}, {Sugimori}, {Suto}, {Suzuki}, {Takada}, {Takagi},
  {Takata}, {Takita}, {Tanaka}, {Tang}, {Taranu}, {Terai}, {Toba}, {Turner},
  {Uchiyama}, {Vijarnwannaluk}, {Waters}, {Yamada}, {Yamamoto}, \&
  {Yamashita}}]{HSC_DR3}
{Aihara}, H., {AlSayyad}, Y., {Ando}, M., {et~al.} 2022, \pasj, 74, 247,
  \dodoi{10.1093/pasj/psab122}

\bibitem[{{Argudo-Fern{\'a}ndez} {et~al.}(2015){Argudo-Fern{\'a}ndez},
  {Verley}, {Bergond}, {Duarte Puertas}, {Ramos Carmona}, {Sabater},
  {Fern{\'a}ndez Lorenzo}, {Espada}, {Sulentic}, {Ruiz}, \&
  {Leon}}]{Argudo_2015}
{Argudo-Fern{\'a}ndez}, M., {Verley}, S., {Bergond}, G., {et~al.} 2015, \aap,
  578, A110, \dodoi{10.1051/0004-6361/201526016}

\bibitem[{{Baldry} {et~al.}(2010){Baldry}, {Robotham}, {Hill}, {Driver},
  {Liske}, {Norberg}, {Bamford}, {Hopkins}, {Loveday}, {Peacock}, {Cameron},
  {Croom}, {Cross}, {Doyle}, {Dye}, {Frenk}, {Jones}, {van Kampen}, {Kelvin},
  {Nichol}, {Parkinson}, {Popescu}, {Prescott}, {Sharp}, {Sutherland},
  {Thomas}, \& {Tuffs}}]{Baldry_2010}
{Baldry}, I.~K., {Robotham}, A.~S.~G., {Hill}, D.~T., {et~al.} 2010, \mnras,
  404, 86, \dodoi{10.1111/j.1365-2966.2010.16282.x}

\bibitem[{{Blanton} {et~al.}(2011){Blanton}, {Kazin}, {Muna}, {Weaver}, \&
  {Price-Whelan}}]{Blanton2011}
{Blanton}, M.~R., {Kazin}, E., {Muna}, D., {Weaver}, B.~A., \& {Price-Whelan},
  A. 2011, \aj, 142, 31, \dodoi{10.1088/0004-6256/142/1/31}

\bibitem[{{Boquien} {et~al.}(2019){Boquien}, {Burgarella}, {Roehlly}, {Buat},
  {Ciesla}, {Corre}, {Inoue}, \& {Salas}}]{CIGALE_2019}
{Boquien}, M., {Burgarella}, D., {Roehlly}, Y., {et~al.} 2019, \aap, 622, A103,
  \dodoi{10.1051/0004-6361/201834156}

\bibitem[{{Brodie} {et~al.}(2011){Brodie}, {Romanowsky}, {Strader}, \&
  {Forbes}}]{Brodie2011}
{Brodie}, J.~P., {Romanowsky}, A.~J., {Strader}, J., \& {Forbes}, D.~A. 2011,
  \aj, 142, 199, \dodoi{10.1088/0004-6256/142/6/199}

\bibitem[{{Cappellari}(2017)}]{Cappellari_ppxf2017}
{Cappellari}, M. 2017, MNRAS, 466, 798, \dodoi{10.1093/mnras/stw3020}

\bibitem[{{Chilingarian} {et~al.}(2009){Chilingarian}, {Cayatte}, {Revaz},
  {Dodonov}, {Durand}, {Durret}, {Micol}, \& {Slezak}}]{Chilingarian_2009Sci}
{Chilingarian}, I., {Cayatte}, V., {Revaz}, Y., {et~al.} 2009, Science, 326,
  1379, \dodoi{10.1126/science.1175930}

\bibitem[{{Chilingarian} \& {Zolotukhin}(2015)}]{Chilingarian_2015Sci}
{Chilingarian}, I., \& {Zolotukhin}, I. 2015, Science, 348, 418,
  \dodoi{10.1126/science.aaa3344}

\bibitem[{{Chilingarian} \& {Bergond}(2010)}]{Chilingarian_2010}
{Chilingarian}, I.~V., \& {Bergond}, G. 2010, \mnras, 405, L11,
  \dodoi{10.1111/j.1745-3933.2010.00849.x}

\bibitem[{{Diemand} {et~al.}(2005){Diemand}, {Madau}, \& {Moore}}]{Diemand2005}
{Diemand}, J., {Madau}, P., \& {Moore}, B. 2005, \mnras, 364, 367,
  \dodoi{10.1111/j.1365-2966.2005.09604.x}

\bibitem[{{Drinkwater} {et~al.}(2000){Drinkwater}, {Jones}, {Gregg}, \&
  {Phillipps}}]{Drinkwater_2000}
{Drinkwater}, M.~J., {Jones}, J.~B., {Gregg}, M.~D., \& {Phillipps}, S. 2000,
  \pasa, 17, 227, \dodoi{10.1071/AS00034}

\bibitem[{{Driver} {et~al.}(2011){Driver}, {Hill}, {Kelvin}, {Robotham},
  {Liske}, {Norberg}, {Baldry}, {Bamford}, {Hopkins}, {Loveday}, {Peacock},
  {Andrae}, {Bland-Hawthorn}, {Brough}, {Brown}, {Cameron}, {Ching}, {Colless},
  {Conselice}, {Croom}, {Cross}, {de Propris}, {Dye}, {Drinkwater}, {Ellis},
  {Graham}, {Grootes}, {Gunawardhana}, {Jones}, {van Kampen}, {Maraston},
  {Nichol}, {Parkinson}, {Phillipps}, {Pimbblet}, {Popescu}, {Prescott},
  {Roseboom}, {Sadler}, {Sansom}, {Sharp}, {Smith}, {Taylor}, {Thomas},
  {Tuffs}, {Wijesinghe}, {Dunne}, {Frenk}, {Jarvis}, {Madore}, {Meyer},
  {Seibert}, {Staveley-Smith}, {Sutherland}, \& {Warren}}]{Driver_2011_GAMA}
{Driver}, S.~P., {Hill}, D.~T., {Kelvin}, L.~S., {et~al.} 2011, \mnras, 413,
  971, \dodoi{10.1111/j.1365-2966.2010.18188.x}

\bibitem[{{Driver} {et~al.}(2022){Driver}, {Bellstedt}, {Robotham}, {Baldry},
  {Davies}, {Liske}, {Obreschkow}, {Taylor}, {Wright}, {Alpaslan}, {Bamford},
  {Bauer}, {Bland-Hawthorn}, {Bilicki}, {Bravo}, {Brough}, {Casura}, {Cluver},
  {Colless}, {Conselice}, {Croom}, {de Jong}, {D'Eugenio}, {De Propris},
  {Dogruel}, {Drinkwater}, {Dvornik}, {Farrow}, {Frenk}, {Giblin}, {Graham},
  {Grootes}, {Gunawardhana}, {Hashemizadeh}, {H{\"a}u{\ss}ler}, {Heymans},
  {Hildebrandt}, {Holwerda}, {Hopkins}, {Jarrett}, {Heath Jones}, {Kelvin},
  {Koushan}, {Kuijken}, {Lara-L{\'o}pez}, {Lange}, {L{\'o}pez-S{\'a}nchez},
  {Loveday}, {Mahajan}, {Meyer}, {Moffett}, {Napolitano}, {Norberg}, {Owers},
  {Radovich}, {Raouf}, {Peacock}, {Phillipps}, {Pimbblet}, {Popescu}, {Said},
  {Sansom}, {Seibert}, {Sutherland}, {Thorne}, {Tuffs}, {Turner}, {van der
  Wel}, {van Kampen}, \& {Wilkins}}]{GAMA_DR4}
{Driver}, S.~P., {Bellstedt}, S., {Robotham}, A. S.~G., {et~al.} 2022, \mnras,
  513, 439, \dodoi{10.1093/mnras/stac472}

\bibitem[{{Du} {et~al.}(2019){Du}, {Debattista}, {Ho}, {C{\^o}t{\'e}},
  {Spengler}, {Erwin}, {Wadsley}, {Norris}, {Earp}, {Quinn}, {Fiteni}, \&
  {Caruana}}]{Du2019}
{Du}, M., {Debattista}, V.~P., {Ho}, L.~C., {et~al.} 2019, \apj, 875, 58,
  \dodoi{10.3847/1538-4357/ab0e0c}

\bibitem[{{Eigenthaler} {et~al.}(2018){Eigenthaler}, {Puzia}, {Taylor},
  {Ordenes-Brice{\~n}o}, {Mu{\~n}oz}, {Ribbeck}, {Alamo-Mart{\'\i}nez},
  {Zhang}, {{\'A}ngel}, {Capaccioli}, {C{\^o}t{\'e}}, {Ferrarese}, {Galaz},
  {Grebel}, {Hempel}, {Hilker}, {Lan{\c{c}}on}, {Mieske}, {Miller}, {Paolillo},
  {Powalka}, {Richtler}, {Roediger}, {Rong}, {S{\'a}nchez-Janssen}, \&
  {Spengler}}]{Eigenthaler_2018}
{Eigenthaler}, P., {Puzia}, T.~H., {Taylor}, M.~A., {et~al.} 2018, \apj, 855,
  142, \dodoi{10.3847/1538-4357/aaab60}

\bibitem[{{Fellhauer} \& {Kroupa}(2002)}]{Fellhauer_2002}
{Fellhauer}, M., \& {Kroupa}, P. 2002, \mnras, 330, 642,
  \dodoi{10.1046/j.1365-8711.2002.05087.x}

\bibitem[{{Ferr{\'e}-Mateu} {et~al.}(2018){Ferr{\'e}-Mateu}, {Forbes},
  {Romanowsky}, {Janz}, \& {Dixon}}]{Ferre_2018}
{Ferr{\'e}-Mateu}, A., {Forbes}, D.~A., {Romanowsky}, A.~J., {Janz}, J., \&
  {Dixon}, C. 2018, \mnras, 473, 1819, \dodoi{10.1093/mnras/stx2442}

\bibitem[{{Gu{\'e}rou} {et~al.}(2015){Gu{\'e}rou}, {Emsellem}, {McDermid},
  {C{\^o}t{\'e}}, {Ferrarese}, {Blakeslee}, {Durrell}, {MacArthur}, {Peng},
  {Cuillandre}, \& {Gwyn}}]{Guerou_2015}
{Gu{\'e}rou}, A., {Emsellem}, E., {McDermid}, R.~M., {et~al.} 2015, \apj, 804,
  70, \dodoi{10.1088/0004-637X/804/1/70}

\bibitem[{{Hilker} {et~al.}(1999){Hilker}, {Infante}, \&
  {Richtler}}]{Hilker_1999}
{Hilker}, M., {Infante}, L., \& {Richtler}, T. 1999, \aaps, 138, 55,
  \dodoi{10.1051/aas:1999495}

\bibitem[{{Hopkins} {et~al.}(2010){Hopkins}, {Murray}, {Quataert}, \&
  {Thompson}}]{Hopkins2010}
{Hopkins}, P.~F., {Murray}, N., {Quataert}, E., \& {Thompson}, T.~A. 2010,
  \mnras, 401, L19, \dodoi{10.1111/j.1745-3933.2009.00777.x}

\bibitem[{{Huxor} {et~al.}(2013){Huxor}, {Phillipps}, \& {Price}}]{Huxor_2013}
{Huxor}, A.~P., {Phillipps}, S., \& {Price}, J. 2013, \mnras, 430, 1956,
  \dodoi{10.1093/mnras/stt014}

\bibitem[{{Huxor} {et~al.}(2011){Huxor}, {Phillipps}, {Price}, \&
  {Harniman}}]{Huxor_2011}
{Huxor}, A.~P., {Phillipps}, S., {Price}, J., \& {Harniman}, R. 2011, \mnras,
  414, 3557, \dodoi{10.1111/j.1365-2966.2011.18655.x}

\bibitem[{{Janz} {et~al.}(2016){Janz}, {Norris}, {Forbes}, {Huxor},
  {Romanowsky}, {Frank}, {Escudero}, {Faifer}, {Forte}, {Kannappan},
  {Maraston}, {Brodie}, {Strader}, \& {Thompson}}]{Janz_2016}
{Janz}, J., {Norris}, M.~A., {Forbes}, D.~A., {et~al.} 2016, \mnras, 456, 617,
  \dodoi{10.1093/mnras/stv2636}

\bibitem[{{Kennicutt} \& {Evans}(2012)}]{Kennicutt_2012}
{Kennicutt}, R.~C., \& {Evans}, N.~J. 2012, \araa, 50, 531,
  \dodoi{10.1146/annurev-astro-081811-125610}

\bibitem[{{Kim} {et~al.}(2020){Kim}, {Jeong}, {Rey}, {Lee}, {Lee}, {Joo}, \&
  {Kim}}]{Kim_2020}
{Kim}, S., {Jeong}, H., {Rey}, S.-C., {et~al.} 2020, \apj, 903, 65,
  \dodoi{10.3847/1538-4357/abaef5}

\bibitem[{{Kravtsov}(2013)}]{Kravtsov_2013}
{Kravtsov}, A.~V. 2013, \apjl, 764, L31, \dodoi{10.1088/2041-8205/764/2/L31}

\bibitem[{{Liske} {et~al.}(2015){Liske}, {Baldry}, {Driver}, {Tuffs},
  {Alpaslan}, {Andrae}, {Brough}, {Cluver}, {Grootes}, {Gunawardhana},
  {Kelvin}, {Loveday}, {Robotham}, {Taylor}, {Bamford}, {Bland-Hawthorn},
  {Brown}, {Drinkwater}, {Hopkins}, {Meyer}, {Norberg}, {Peacock}, {Agius},
  {Andrews}, {Bauer}, {Ching}, {Colless}, {Conselice}, {Croom}, {Davies}, {De
  Propris}, {Dunne}, {Eardley}, {Ellis}, {Foster}, {Frenk}, {H{\"a}u{\ss}ler},
  {Holwerda}, {Howlett}, {Ibarra}, {Jarvis}, {Jones}, {Kafle}, {Lacey},
  {Lange}, {Lara-L{\'o}pez}, {L{\'o}pez-S{\'a}nchez}, {Maddox}, {Madore},
  {McNaught-Roberts}, {Moffett}, {Nichol}, {Owers}, {Palamara}, {Penny},
  {Phillipps}, {Pimbblet}, {Popescu}, {Prescott}, {Proctor}, {Sadler},
  {Sansom}, {Seibert}, {Sharp}, {Sutherland}, {V{\'a}zquez-Mata}, {van Kampen},
  {Wilkins}, {Williams}, \& {Wright}}]{Liske_2015}
{Liske}, J., {Baldry}, I.~K., {Driver}, S.~P., {et~al.} 2015, \mnras, 452,
  2087, \dodoi{10.1093/mnras/stv1436}

\bibitem[{{Liu} {et~al.}(2015{\natexlab{a}}){Liu}, {Peng}, {C{\^o}t{\'e}},
  {Ferrarese}, {Jord{\'a}n}, {Mihos}, {Zhang}, {Mu{\~n}oz}, {Puzia},
  {Lan{\c{c}}on}, {Gwyn}, {Cuillandre}, {Blakeslee}, {Boselli}, {Durrell},
  {Duc}, {Guhathakurta}, {MacArthur}, {Mei}, {S{\'a}nchez-Janssen}, \&
  {Xu}}]{Liu_2015a}
{Liu}, C., {Peng}, E.~W., {C{\^o}t{\'e}}, P., {et~al.} 2015{\natexlab{a}},
  \apj, 812, 34, \dodoi{10.1088/0004-637X/812/1/34}

\bibitem[{{Liu} {et~al.}(2015{\natexlab{b}}){Liu}, {Peng}, {Toloba}, {Mihos},
  {Ferrarese}, {Alamo-Mart{\'\i}nez}, {Zhang}, {C{\^o}t{\'e}}, {Cuillandre},
  {Cunningham}, {Guhathakurta}, {Gwyn}, {Herczeg}, {Lim}, {Puzia}, {Roediger},
  {S{\'a}nchez-Janssen}, \& {Yin}}]{Liu_2015b}
{Liu}, C., {Peng}, E.~W., {Toloba}, E., {et~al.} 2015{\natexlab{b}}, \apjl,
  812, L2, \dodoi{10.1088/2041-8205/812/1/L2}

\bibitem[{{Liu} {et~al.}(2020){Liu}, {C{\^o}t{\'e}}, {Peng}, {Roediger},
  {Zhang}, {Ferrarese}, {S{\'a}nchez-Janssen}, {Guhathakurta}, {Yang}, {Jing},
  {Alamo-Mart{\'\i}nez}, {Blakeslee}, {Boselli}, {Cuilandre}, {Duc}, {Durrell},
  {Gwyn}, {Jord{\'a}n}, {Ko}, {Lan{\c{c}}on}, {Lim}, {Longobardi}, {Mei},
  {Mihos}, {Mu{\~n}oz}, {Powalka}, {Puzia}, {Spengler}, \& {Toloba}}]{Liu_2020}
{Liu}, C., {C{\^o}t{\'e}}, P., {Peng}, E.~W., {et~al.} 2020, \apjs, 250, 17,
  \dodoi{10.3847/1538-4365/abad91}

\bibitem[{{Mayes} {et~al.}(2021){Mayes}, {Drinkwater}, {Pfeffer}, {Baumgardt},
  {Liu}, {Ferrarese}, {C{\^o}t{\'e}}, \& {Peng}}]{Mayes2021}
{Mayes}, R.~J., {Drinkwater}, M.~J., {Pfeffer}, J., {et~al.} 2021, \mnras, 501,
  1852, \dodoi{10.1093/mnras/staa3731}

\bibitem[{{Mieske} {et~al.}(2012){Mieske}, {Hilker}, \& {Misgeld}}]{Mieske2012}
{Mieske}, S., {Hilker}, M., \& {Misgeld}, I. 2012, \aap, 537, A3,
  \dodoi{10.1051/0004-6361/201117634}

\bibitem[{{Misgeld} \& {Hilker}(2011)}]{Misgeld2011}
{Misgeld}, I., \& {Hilker}, M. 2011, \mnras, 414, 3699,
  \dodoi{10.1111/j.1365-2966.2011.18669.x}

\bibitem[{{Norris} {et~al.}(2014){Norris}, {Kannappan}, {Forbes}, {Romanowsky},
  {Brodie}, {Faifer}, {Huxor}, {Maraston}, {Moffett}, {Penny}, {Pota},
  {Smith-Castelli}, {Strader}, {Bradley}, {Eckert}, {Fohring}, {McBride},
  {Stark}, \& {Vaduvescu}}]{Norris2014}
{Norris}, M.~A., {Kannappan}, S.~J., {Forbes}, D.~A., {et~al.} 2014, \mnras,
  443, 1151, \dodoi{10.1093/mnras/stu1186}

\bibitem[{{Paudel} {et~al.}(2014){Paudel}, {Lisker}, {Hansson}, \&
  {Huxor}}]{Paudel_2014}
{Paudel}, S., {Lisker}, T., {Hansson}, K.~S.~A., \& {Huxor}, A.~P. 2014,
  \mnras, 443, 446, \dodoi{10.1093/mnras/stu1171}

\bibitem[{{Pechetti} {et~al.}(2022){Pechetti}, {Seth}, {Kamann}, {Caldwell},
  {Strader}, {den Brok}, {Luetzgendorf}, {Neumayer}, \&
  {Voggel}}]{Pechetti2022}
{Pechetti}, R., {Seth}, A., {Kamann}, S., {et~al.} 2022, \apj, 924, 48,
  \dodoi{10.3847/1538-4357/ac339f}

\bibitem[{{Peng} {et~al.}(2010){Peng}, {Ho}, {Impey}, \& {Rix}}]{Peng_2010}
{Peng}, C.~Y., {Ho}, L.~C., {Impey}, C.~D., \& {Rix}, H.-W. 2010, \aj, 139,
  2097, \dodoi{10.1088/0004-6256/139/6/2097}

\bibitem[{{Pfeffer} \& {Baumgardt}(2013)}]{Pfeffer2013}
{Pfeffer}, J., \& {Baumgardt}, H. 2013, \mnras, 433, 1997,
  \dodoi{10.1093/mnras/stt867}

\bibitem[{{Price} {et~al.}(2009){Price}, {Phillipps}, {Huxor}, {Trentham},
  {Ferguson}, {Marzke}, {Hornschemeier}, {Goudfrooij}, {Hammer}, {Tully},
  {Chiboucas}, {Smith}, {Carter}, {Merritt}, {Balcells}, {Erwin}, \&
  {Puzia}}]{Price2009}
{Price}, J., {Phillipps}, S., {Huxor}, A., {et~al.} 2009, \mnras, 397, 1816,
  \dodoi{10.1111/j.1365-2966.2009.15122.x}

\bibitem[{{Rey} {et~al.}(2021){Rey}, {Oh}, \& {Kim}}]{Rey_2021}
{Rey}, S.-C., {Oh}, K., \& {Kim}, S. 2021, \apjl, 917, L9,
  \dodoi{10.3847/2041-8213/ac15f6}

\bibitem[{{Robotham} {et~al.}(2011){Robotham}, {Norberg}, {Driver}, {Baldry},
  {Bamford}, {Hopkins}, {Liske}, {Loveday}, {Merson}, {Peacock}, {Brough},
  {Cameron}, {Conselice}, {Croom}, {Frenk}, {Gunawardhana}, {Hill}, {Jones},
  {Kelvin}, {Kuijken}, {Nichol}, {Parkinson}, {Pimbblet}, {Phillipps},
  {Popescu}, {Prescott}, {Sharp}, {Sutherland}, {Taylor}, {Thomas}, {Tuffs},
  {van Kampen}, \& {Wijesinghe}}]{GAMA_G3C_2011}
{Robotham}, A.~S.~G., {Norberg}, P., {Driver}, S.~P., {et~al.} 2011, \mnras,
  416, 2640, \dodoi{10.1111/j.1365-2966.2011.19217.x}

\bibitem[{{Saifollahi} {et~al.}(2021){Saifollahi}, {Janz}, {Peletier},
  {Cantiello}, {Hilker}, {Mieske}, {Valentijn}, {Venhola}, \&
  {Kleijn}}]{Saifollahi2021}
{Saifollahi}, T., {Janz}, J., {Peletier}, R.~F., {et~al.} 2021, \mnras, 504,
  3580, \dodoi{10.1093/mnras/stab1118}

\bibitem[{{Schlegel} {et~al.}(1998){Schlegel}, {Finkbeiner}, \&
  {Davis}}]{Schlegel1998}
{Schlegel}, D.~J., {Finkbeiner}, D.~P., \& {Davis}, M. 1998, \apj, 500, 525,
  \dodoi{10.1086/305772}

\bibitem[{{Seth} {et~al.}(2014){Seth}, {van den Bosch}, {Mieske}, {Baumgardt},
  {Brok}, {Strader}, {Neumayer}, {Chilingarian}, {Hilker}, {McDermid},
  {Spitler}, {Brodie}, {Frank}, \& {Walsh}}]{Seth_2014}
{Seth}, A.~C., {van den Bosch}, R., {Mieske}, S., {et~al.} 2014, \nat, 513,
  398, \dodoi{10.1038/nature13762}

\bibitem[{{Strader} {et~al.}(2013){Strader}, {Seth}, {Forbes}, {Fabbiano},
  {Romanowsky}, {Brodie}, {Conroy}, {Caldwell}, {Pota}, {Usher}, \&
  {Arnold}}]{Strader_2013}
{Strader}, J., {Seth}, A.~C., {Forbes}, D.~A., {et~al.} 2013, \apjl, 775, L6,
  \dodoi{10.1088/2041-8205/775/1/L6}

\bibitem[{{Weinmann} {et~al.}(2006){Weinmann}, {van den Bosch}, {Yang}, \&
  {Mo}}]{Weinmann2006}
{Weinmann}, S.~M., {van den Bosch}, F.~C., {Yang}, X., \& {Mo}, H.~J. 2006,
  \mnras, 366, 2, \dodoi{10.1111/j.1365-2966.2005.09865.x}

\bibitem[{{Wright} {et~al.}(2010){Wright}, {Eisenhardt}, {Mainzer}, {Ressler},
  {Cutri}, {Jarrett}, {Kirkpatrick}, {Padgett}, {McMillan}, {Skrutskie},
  {Stanford}, {Cohen}, {Walker}, {Mather}, {Leisawitz}, {Gautier}, {McLean},
  {Benford}, {Lonsdale}, {Blain}, {Mendez}, {Irace}, {Duval}, {Liu}, {Royer},
  {Heinrichsen}, {Howard}, {Shannon}, {Kendall}, {Walsh}, {Larsen}, {Cardon},
  {Schick}, {Schwalm}, {Abid}, {Fabinsky}, {Naes}, \& {Tsai}}]{Wright_2010}
{Wright}, E.~L., {Eisenhardt}, P. R.~M., {Mainzer}, A.~K., {et~al.} 2010, \aj,
  140, 1868, \dodoi{10.1088/0004-6256/140/6/1868}

\bibitem[{{York} {et~al.}(2000){York}, {Adelman}, {Anderson}, {Anderson},
  {Annis}, {Bahcall}, {Bakken}, {Barkhouser}, {Bastian}, {Berman}, {Boroski},
  {Bracker}, {Briegel}, {Briggs}, {Brinkmann}, {Brunner}, {Burles}, {Carey},
  {Carr}, {Castander}, {Chen}, {Colestock}, {Connolly}, {Crocker}, {Csabai},
  {Czarapata}, {Davis}, {Doi}, {Dombeck}, {Eisenstein}, {Ellman}, {Elms},
  {Evans}, {Fan}, {Federwitz}, {Fiscelli}, {Friedman}, {Frieman}, {Fukugita},
  {Gillespie}, {Gunn}, {Gurbani}, {de Haas}, {Haldeman}, {Harris}, {Hayes},
  {Heckman}, {Hennessy}, {Hindsley}, {Holm}, {Holmgren}, {Huang}, {Hull},
  {Husby}, {Ichikawa}, {Ichikawa}, {Ivezi{\'c}}, {Kent}, {Kim}, {Kinney},
  {Klaene}, {Kleinman}, {Kleinman}, {Knapp}, {Korienek}, {Kron}, {Kunszt},
  {Lamb}, {Lee}, {Leger}, {Limmongkol}, {Lindenmeyer}, {Long}, {Loomis},
  {Loveday}, {Lucinio}, {Lupton}, {MacKinnon}, {Mannery}, {Mantsch}, {Margon},
  {McGehee}, {McKay}, {Meiksin}, {Merelli}, {Monet}, {Munn}, {Narayanan},
  {Nash}, {Neilsen}, {Neswold}, {Newberg}, {Nichol}, {Nicinski}, {Nonino},
  {Okada}, {Okamura}, {Ostriker}, {Owen}, {Pauls}, {Peoples}, {Peterson},
  {Petravick}, {Pier}, {Pope}, {Pordes}, {Prosapio}, {Rechenmacher}, {Quinn},
  {Richards}, {Richmond}, {Rivetta}, {Rockosi}, {Ruthmansdorfer}, {Sandford},
  {Schlegel}, {Schneider}, {Sekiguchi}, {Sergey}, {Shimasaku}, {Siegmund},
  {Smee}, {Smith}, {Snedden}, {Stone}, {Stoughton}, {Strauss}, {Stubbs},
  {SubbaRao}, {Szalay}, {Szapudi}, {Szokoly}, {Thakar}, {Tremonti}, {Tucker},
  {Uomoto}, {Vanden Berk}, {Vogeley}, {Waddell}, {Wang}, {Watanabe},
  {Weinberg}, {Yanny}, {Yasuda}, \& {SDSS Collaboration}}]{York_2000}
{York}, D.~G., {Adelman}, J., {Anderson}, John~E., J., {et~al.} 2000, \aj, 120,
  1579, \dodoi{10.1086/301513}

\bibitem[{{Zhang} {et~al.}(2015){Zhang}, {Peng}, {C{\^o}t{\'e}}, {Liu},
  {Ferrarese}, {Cuillandre}, {Caldwell}, {Gwyn}, {Jord{\'a}n}, {Lan{\c{c}}on},
  {Li}, {Mu{\~n}oz}, {Puzia}, {Bekki}, {Blakeslee}, {Boselli}, {Drinkwater},
  {Duc}, {Durrell}, {Emsellem}, {Firth}, \& {S{\'a}nchez-Janssen}}]{Zhang_2015}
{Zhang}, H.-X., {Peng}, E.~W., {C{\^o}t{\'e}}, P., {et~al.} 2015, \apj, 802,
  30, \dodoi{10.1088/0004-637X/802/1/30}

\bibitem[{{Zhang} {et~al.}(2018){Zhang}, {Puzia}, {Peng}, {Liu},
  {C{\^o}t{\'e}}, {Ferrarese}, {Duc}, {Eigenthaler}, {Lim}, {Lan{\c{c}}on},
  {Mu{\~n}oz}, {Roediger}, {S{\'a}nchez-Janssen}, {Taylor}, \&
  {Yu}}]{Zhang_2018}
{Zhang}, H.-X., {Puzia}, T.~H., {Peng}, E.~W., {et~al.} 2018, \apj, 858, 37,
  \dodoi{10.3847/1538-4357/aab88a}

\end{thebibliography}
\bibliographystyle{aasjournal}



\end{document}